\documentclass[12pt]{article}
\title{ The Lattice Equations of the Toda Type with an Interaction between a 
Few Neighborhoods }
\author{N.V. Ustinov\\
\\
\it \small Theoretical Physics Department, Kaliningrad State University,\\
\it \small Al.\,Nevsky street 14, 236041, Kaliningrad, Russia}
\date{ }
\begin{document}
\maketitle
\begin{abstract}
The sets of the integrable lattice equations, which generalize the Toda 
lattice, are considered.
The hierarchies of the first integrals and infinitesimal symmetries are found. 
The properties of the multi-soliton solutions are discussed.
\end{abstract}



\section{Introduction}

The first differential--difference system studied by the methods of the theory 
of solitons was the famous Toda lattice \cite{T}
\begin{equation}
\ddot\sigma_k=\mbox{e}^{\displaystyle\sigma_{k+1}-\sigma_k}
-\mbox{e}^{\displaystyle\sigma_k-\sigma_{k-1}},
\label{Tl}
\end{equation}
whose multi-soliton solutions were built using the Hirota method \cite{H}.
The complete integrability of the Toda lattice in periodic and infinitely 
dimensional cases was proven in works \cite{F} and \cite{M}.
This system was originally suggested as one describing the behaviour of the 
particles interacting with the nearest neighborhoods. 
Nevertheless, it appears as a limit in the elliptic Calogero--Moser system  
\cite{In}, where all the particles interact with each other. 
The development of the theory of solitons led to various generalizations of 
the Toda lattice (see, e.g., \cite{Per}). 
The problem of the classification of the integrable lattices of the Toda type 
with an interaction between two nearest neighborhoods was considered in 
\cite{Ya,ASh}. 
Recently, the new classes of the integrable lattice equations that include the 
Toda lattice as the particular case \cite{Bl,Sv,CCzU} and the nonlocal 
two--dimensional generalization of Eqs.(\ref{Tl}) \cite{U} were investigated. 

A significant attention was paid on the integrable multi-particle systems 
during the past ten years. 
It was shown that Eqs.(\ref{Tl}) and other well-known lattice equations 
are the discrete symmetries (Darboux--B\"acklund transformations) of the 
Kadomtsev--Petviashvilii hierarchies \cite{AFGZ,LShYa}. 
Also, these systems play an important role in nonperturbative string theory 
and D-branes theory \cite{Marsh}. 
In particular, the Toda lattice is connected with the Nahm equations 
\cite{Nahm,Hitchin} and determines the behaviour of the collective coordinates 
of the branes \cite{D,GGM} and massless supersymmetric gauge theories in 
low-energy sector \cite{GKMMM,MW,IM}. 

This article is devoted to the generalization of Eqs.(\ref{Tl}) on the cases 
of the systems of particles, whose motion is immediately affected by the 
finite number of nearest neighborhoods. 
These lattices are the members of the one-parameter subvarieties of two 
different sets of the two-parameter two-field integrable lattice equations 
presented in \cite{Sv} and \cite{CCzU}.
It is remarkable that, as for the Toda lattice case, there exists a connection 
of the lattices considered with the Nahm equations and their continuous limit 
is the Boussinesque equation. 
The reductions of the lattice equations, the Lax pairs, the asymptotic 
expansions of the solutions of the Lax pairs and the first integrals are 
given in Sec.II. 
The Darboux transformation technique \cite{MaSa} is applied in Sec.III to 
obtain the hierarchies of the infinitesimal symmetries and the soliton 
solutions.

\section{Lattice Equations of the Toda Type}

Let us consider two infinite sets ($l\in\bf N$) of the lattice equations 
\begin{equation}
\begin{array}{c}
\ddot\sigma_k=C\Bigl(
\mbox{e}^{\sum_{m=0}^{l-1}\displaystyle(\sigma_{k+m+l}-\sigma_{k-m})}
-\mbox{e}^{\sum_{m=0}^{l-1}\displaystyle(\sigma_{k+m}-\sigma_{k-l-m})}
\Bigr)+\mbox{}\\
\mbox{}+\dot\sigma_k\displaystyle\sum_{m=1}^{l-1}(\dot\sigma_{k+m}-\dot\sigma_{k-m})
\end{array}
\label{s_1}
\end{equation}
and 
\begin{equation}
\begin{array}{c}
\ddot\sigma_k=C\Bigl(
\mbox{e}^{\sum_{m=1}^l\displaystyle(\sigma_{k-m}-\sigma_{k+l+m})}
-\mbox{e}^{\sum_{m=1}^l\displaystyle(\sigma_{k-l-m}-\sigma_{k+m})}
\Bigr)-\mbox{}\\
\mbox{}-\dot\sigma_k\displaystyle\sum_{m=1}^l(\dot\sigma_{k+m}-\dot\sigma_{k-m}).
\end{array}
\label{s_2}
\end{equation}
Here $C$ is arbitrary constant, which is assumed to be unequal to zero. 
(If $C=0$, then Eqs.(\ref{s_1},\ref{s_2}) are the Bogoyavlenskii lattice 
\cite{B1}.)
Eqs.(\ref{s_1}) with $l=1$ are evidently reduced to the Toda lattice 
(\ref{Tl}). 
The Belov--Chaltikian lattice \cite{BCh} is equivalent to Eqs.(\ref{s_2}) if 
$l=1$.

In the periodic case ($\sigma_{k+n}=\sigma_k$, $n$ is a period), the lattices 
admit additional reduction constraints\,:
$$
\sigma_{m+1+k}=-\sigma_{m+1-k}\quad\mbox{ if }\quad n=2m+1,
$$
$$
\sigma_{m+k}=-\sigma_{m+1-k}\quad\mbox{ if }\quad n=2m
$$
or
$$
\sigma_{m+k}=-\sigma_{m-k}\quad\mbox{ if }\quad n=2m
$$
($k=0,...,m$). 
A connection of these constraints in the Toda lattice case with the root 
systems of semisimple Lie algebras was established in \cite{B2}.
Reductions 
$$
\sigma_{-k}=-\sigma_k
$$
or
$$
\sigma_{1-k}=-\sigma_k
$$
are consistent with the lattice equations in infinitely dimensional case.

It is well known that the Boussinesque equation 
\begin{equation}
v_{\tau\tau}=(a^2v+c_2v_{xx}+c_3v^2)_{xx}
\label{Busq}
\end{equation}
can be obtained as a result of the limiting procedure in the Toda lattice 
\cite{AS}.
This procedure is suitable in Eqs.(\ref{s_1},\ref{s_2}) for arbitrary $l$ and 
gives the same equation.  
Indeed, assuming 
$$
\sigma_k=\varepsilon u(\tau, x),\quad\tau=\varepsilon^2t,\quad 
x=\varepsilon k
$$
and expanding $u(\tau, x+\varepsilon m)$ in the Taylor series, we have from 
Eqs.(\ref{s_1},\ref{s_2})
$$
u_{\tau\tau}=\frac{c_1}{\varepsilon^2}u_{xx}+c_2u_{xxxx}+2c_3u_xu_{xx}+
O(\varepsilon). 
$$
Eq.(\ref{Busq}) follows this equality after differentiation if we put 
$$
v=u_x+\frac{c_1-\varepsilon^2a^2}{2c_3\varepsilon^2}
$$
and consider limit $\varepsilon\to 0$.

Eqs.(\ref{s_1},\ref{s_2}) with arbitrary $l$ were revealed in \cite{CCzU} to 
belong a class of nonlinear equations representable as the compatibility 
condition of overdetermined linear systems (Lax pairs), whose coefficients are 
explicitly connected.
Thus, Lax pair for Eqs.(\ref{s_1}) has form
\begin{equation}
\left\{
\begin{array}{rcl}
-\dot\psi_k&=&\displaystyle\lambda\psi_{k-l}+\sum_{m=0}^{l-1}h_{k+m}\psi_k\\
z\psi_k&=&\lambda^2\psi_{k-1}+\lambda h_{k+l-1}\psi_{k+l-1}+
\rho_{k+2l-1}\psi_{k+2l-1}
\end{array}
\right.,
\label{Lax_1}
\end{equation}
where $z$ and $\lambda$ are complex parameters and 
$$
h_k=\dot\sigma_k,
$$
$$
\rho_k=C\mbox{e}^{\sum_{m=0}^{l-1}\displaystyle(\sigma_{k+m}-\sigma_{k-l-m})}.
$$
Also, this lattice is the compatibility condition of so-called dual Lax pair 
\begin{equation}
\left\{
\begin{array}{rcl}
\dot\xi_k&=&\displaystyle\lambda\xi_{k+l}+\sum_{m=0}^{l-1}h_{k+m}\xi_k\\
z\xi_k&=&\lambda^2\xi_{k+1}+\lambda h_{k}\xi_{k-l+1}+\rho_{k}\xi_{k-2l+1}
\end{array}
\right..
\label{Lax_1d}
\end{equation}
Direct and dual Lax pairs of Eqs.(\ref{s_2}) read respectively as 
\begin{equation}
\left\{
\begin{array}{rcl}
-\dot\psi_k&=&\displaystyle\lambda\psi_{k-l}-\sum_{m=1}^lh_{k+m}\psi_k\\
z\psi_k&=&\lambda^2\psi_{k+1}+\lambda h_{k+l+1}\psi_{k+l+1}+
\rho_{k+2l+1}\psi_{k+2l+1}
\end{array}
\right.
\label{Lax_2}
\end{equation}
and
\begin{equation}
\left\{
\begin{array}{rcl}
\dot\xi_k&=&\displaystyle\lambda\xi_{k+l}-\sum_{m=1}^lh_{k+m}\xi_k\\
z\xi_k&=&\lambda^2\xi_{k-1}+\lambda h_{k}\xi_{k-l-1}+\rho_{k}\xi_{k-2l-1}
\end{array}
\right..
\label{Lax_2d}
\end{equation}
Here
$$
\rho_k=C\mbox{e}^{\sum_{m=1}^l\displaystyle(\sigma_{k-l-m}-\sigma_{k+m})}.
$$
It is worth to note that the dependence on $\lambda$ of the Lax pairs 
(\ref{Lax_1}--\ref{Lax_2d}) is the same as in the case of the Nahm equations. 
Consequently, lattices (\ref{s_1},\ref{s_2}) can be obtained from the Nahm 
equations by imposing the reduction constraints. 

One can put $\lambda=1$ in infinitely dimensional case without loss of 
generality.
Then the solutions of Lax pairs (\ref{Lax_1},\ref{Lax_1d}) admit in the 
neighborhood of the point $z=\infty$ the following asymptotic expansions: 
\begin{equation}
\psi_k=\alpha^k\mbox{e}^{-t/\alpha^l}\left(1+\frac{a_k}{z^l}+
\frac{b_k}{z^{2l}}+\dots\right),
\label{psi_i}
\end{equation}
\begin{equation}
\xi_k=\alpha^{-k}\mbox{e}^{t/\alpha^l}\left(1-\frac{a_{k-l}}{z^l}+
\frac{c_k}{z^{2l}}+\dots\right),
\label{xi_i}
\end{equation}
where
$$
a_k=\sum_{m=-\infty}^{k+l-1}h_m,
$$
$$
b_k=\sum_{m=-\infty}^{k+2l-1}(\rho_m-C+h_{m-l}a_{m-l}),\quad
c_k=\sum_{m=-\infty}^{k-1}(C-\rho_m+h_ma_{m-2l+1})
$$
and 
$$
\alpha=\frac1z\left(1+\frac{C}{z^{2l}}+\dots\right)
$$
is the solution of equation
$$
z\alpha=1+C\alpha^{2l}.
$$
The solutions of Lax pairs (\ref{Lax_2},\ref{Lax_2d}) can be represented in 
the next form 
\begin{equation}
\psi_k=\beta^k\mbox{e}^{-t/\beta^l}\left(1+z^ld_k+z^{2l}e_k+\dots\right),
\label{psi_0}
\end{equation}
\begin{equation}
\xi_k=\beta^{-k}\mbox{e}^{t/\beta^l}\left(1-z^ld_{k-l}+z^{2l}f_k+\dots\right)
\label{xi_0}
\end{equation}
in the neighborhood of the point $z=0$.
Here
$$
d_k=-\sum_{m=-\infty}^{k+l}h_m,
$$
$$
e_k=\sum_{m=-\infty}^{k+2l}(C-\rho_m-h_{m-l}d_{m-l}),\quad
f_k=\sum_{m=-\infty}^{k}(\rho_m-C-h_md_{m-2l-1})
$$
and 
$$
\beta=z\left(1-z^{2l}C+\dots\right)
$$
satisfy equation
$$
z=\beta+C\beta^{2l+1}.
$$

The second equations of the Lax pairs (\ref{Lax_1},\ref{Lax_2}) can be 
rewritten as infinitely dimensional spectral problems 
$$
z\psi=L_1\psi
$$
and
$$
z\psi=L_2\psi,
$$
with the help of matrices $L_1$ and $L_2$ such that 
$$
L_{1,kj}=\lambda^2\delta_{k,j+1}+\lambda h_j\delta_{k,j+1-l}+
\rho_j\delta_{k,j+1-2l},
$$
$$
L_{2,kj}=\lambda^2\delta_{k,j-1}+\lambda h_j\delta_{k,j-1-l}+
\rho_j\delta_{k,j-1-2l}.
$$
The quantities 
$$
I_n=\mbox{Tr\,}L_1^{nl}
$$
and 
$$
J_n=\mbox{Tr\,}L_2^{-nl}
$$
($n\in\bf N$) give respectively the infinite hierarchy of the integrals of 
lattices (\ref{s_1}) and (\ref{s_2}). 
The first nontrivial integrals are 
$$
I_2=l\sum_{k=-\infty}^{\infty}\left(2\rho_k+h_k^2+2h_k\sum_{m=1}^{l-1}h_{k+m} 
\right),
$$
$$
J_2=l\sum_{k=-\infty}^{\infty}\left(-2\rho_k+h_k^2+2h_k\sum_{m=1}^lh_{k+m} 
\right).
$$
It is seen that positively defined integral exists only in the case of 
real-valued solutions of the Toda lattice with $C>0$. 

\section{Darboux Transformation Technique}

In this section, we give the formulas of the Darboux transformations (DTs), 
which allow us to generate the infinite hierarchies of the solutions of 
lattices (\ref{s_1}) and (\ref{s_2}) together with ones of their Lax pairs. 
The infinitesimal symmetries of the lattices are also found.

Let $\varphi_k$ be a solution of system (\ref{Lax_1}) with $z=x$ and 
$\lambda=\mu$.
Eqs.(\ref{s_1}) and Lax pairs (\ref{Lax_1},\ref{Lax_1d}) are covariant with 
respect to DT
\begin{equation}
\tilde\psi_k=\dot\psi_k-\frac{\dot\varphi_k}{\varphi_k}\psi_k,
\label{tpsi_1}
\end{equation}
\begin{equation}
\tilde\xi_k=\frac{\Delta_k}{\varphi_{k-l}},
\label{txi_1}
\end{equation}
\begin{equation}
\tilde\sigma_k=\sigma_k+\log\frac{\varphi_{k-l+1}}{\varphi_{k-l}}, 
\label{ts_1}
\end{equation}
where
\begin{equation}
\Delta_k=\int_{t_0}^{t}\varphi_{k-l}\xi_k\,dt+\delta_k, 
\label{D_1}
\end{equation}
constants $\delta_k$ are determined by equations
$$
\lambda\delta_{k+l}-\mu\delta_k=\varphi_k\xi_k|_{t=t_0},
$$
$$
x\lambda^2\delta_{k+1}-z\mu^2\delta_k=\left.\left[
\lambda\mu h_k\varphi_k\xi_{k-l+1}+\lambda\rho_{k+l}\varphi_{k+l}\xi_{k-l+1}+
\mu\rho_k\varphi_k\xi_{k-2l+1}\right]\right|_{t=t_0}.
$$
The statement remains valid, when quantities $\Delta_k$ are defined as
\begin{equation}
\Delta_k=\frac1\mu\sum_{m=1}^{\infty}\left(\frac\mu\lambda\right)^m
\varphi_{k-ml}\xi_{k-ml} 
\label{D_2}
\end{equation}
or
\begin{equation}
\Delta_k=-\frac1\mu\sum_{m=0}^{\infty}\left(\frac\lambda\mu\right)^m
\varphi_{k+ml}\xi_{k+ml} 
\label{D_3}
\end{equation}
We will refer to formulas (\ref{tpsi_1}--\ref{ts_1}) as DT of direct problem 
since the expressions for the transformed quantities depend explicitly on 
solution $\varphi_k$ of direct Lax pair (\ref{Lax_1}). 
If we carry out $N$ iterations of transformation (\ref{tpsi_1}--\ref{ts_1}) on 
solutions $\varphi^{(j)}_k$ ($j=1,...,N$) of Lax pair (\ref{Lax_1}) with 
$z=x^{(j)}$ and $\lambda=\mu^{(j)}$, then new (transformed) solution of 
Eqs.(\ref{s_1}) is 
\begin{equation}
\tilde\sigma_k=\sigma_k+
\log\frac{W(\varphi^{(1)}_{k-l+1},...,\varphi^{(N)}_{k-l+1})}
{W(\varphi^{(1)}_{k-l},...,\varphi^{(N)}_{k-l})}.
\label{ts_1N}
\end{equation}
Here we use notation $W(\varphi^{(1)}_k,...,\varphi^{(N)}_k)$ for the 
Wronskian of functions $\varphi^{(1)}_k,...,\varphi^{(N)}_k$.

In similar manner, one can introduce DT of dual problem with the help of a 
solution of dual Lax pair (\ref{Lax_1d}).
The sum of the DTs of direct and dual problems is called the binary DT and 
yields the following expression for new solution of lattice (\ref{s_1}): 
\begin{equation}
\tilde\sigma_k=\sigma_k+\log\frac{\Delta_{k+1}}{\Delta_k}.
\label{ts_1b}
\end{equation}
Supposing $\lambda=\mu=1$ for the simplicity and considering limit 
$x=z+\delta z\to z$, $\varphi_k=\psi_k+O(\delta z)$ in (\ref{ts_1b}) we obtain 
$$
\tilde\sigma_k=\sigma_k+\delta z\,\delta\sigma_k+o(\delta z),
$$
where
$$
\delta\sigma_k=\Delta_{k+1}-\Delta_k
$$
satisfy the linearization of Eqs.(\ref{s_1}). 
The substitution of expansions (\ref{psi_i},\ref{xi_i}) into the last formula 
leads to the next representation 
$$
\delta\sigma_k=\sum_{m=1}^{\infty}\frac{\delta\sigma_k^{(m)}}{z^{ml}},
$$
where $\delta\sigma_k^{(m)}$ form the infinite hierarchy of the infinitesimal 
symmetries of lattice (\ref{s_1}). 
The first members of the hierarchy are 
$$
\delta\sigma_k^{(1)}=h_k,
$$
$$
\delta\sigma_k^{(2)}=\rho_{k+l}+\rho_k+h_k\sum_{m=-l+1}^{l-1}h_{k+m}. 
$$

Let us consider the zero background ($\sigma_k=0$).
The Lax pairs solutions entering (\ref{ts_1N}) have form  
\begin{equation}
\varphi^{(j)}_k=\sum_{m=1}^{2l}C^{(j)}_m{\alpha^{(j)}_m}^k
\mbox{e}^{\displaystyle-\mu^{(j)}{\alpha^{(j)}_m}^{-l}t},
\label{phi_jk}
\end{equation}
where $\alpha^{(j)}_m$ ($m=1,...,2l$) are the roots of equations 
\begin{equation}
x^{(j)}\alpha^{(j)}={\mu^{(j)}}^2+C{\alpha^{(j)}}^{2l},
\label{a_jk}
\end{equation}
$C^{(j)}_m$ are arbitrary constants.
Substituting (\ref{phi_jk}) into Eqs.(\ref{ts_1N}), we find the multi-soliton 
solution of Eqs.(\ref{s_1}).
(Note that, in the Toda lattice case, relations (\ref{a_jk}) can be 
considered as the equations, which define $x^{(j)}$ and $\mu^{(j)}$ through 
given $\alpha_1^{(j)}$ and $\alpha_2^{(j)}$.)
If all $x^{(j)}$ are different, then the one-soliton component of the 
multi-soliton solution of lattice (\ref{s_1}) is characterized by $2l$ 
parameters: $x^{(j)}$ and $2l-1$ constants from set $C^{(j)}_m$ ($m=1,...2l$). 
These parameters determine the internal degrees of freedom of solitons. 
In the general case, the shape of the one-soliton solution ($N=1$) is changed 
under the evolution. 
If we put 
$$
C^{(j)}_1C^{(j)}_2\ne0,\quad C^{(j)}_m=0\,\,\,(m>2),
$$
then an interaction of the one-soliton components of this subvariety of the 
multi-soliton solutions can lead to their shifts only.
Assuming
$$
C>0,\,\,\,x^{(j)}>0,\,\,\,\mu^{(j)}\in\,{\bf R},
$$
$$
\alpha^{(j)}_1<\alpha^{(j)}_2,\,\,\,v_j>v_k\mbox{ if }j<k,
$$
where
$$
v_j=\mu^{(j)}\frac{{\alpha^{(j)}_1}^{-l}-{\alpha^{(j)}_2}^{-l}}
{\log\left|\alpha^{(j)}_1\right|-\log\left|\alpha^{(j)}_2\right|}
$$
is the velocity of the one-soliton component, we obtain that $j$-th 
one-soliton component suffers shift
$$
\delta_{jk}=\frac{1}{\mu^{(j)}({\alpha^{(j)}_2}^{-l}-{\alpha^{(j)}_1}^{-l})}\log
\frac{(\mu^{(j)}{\alpha^{(j)}_1}^{-l}-\mu^{(k)}{\alpha^{(k)}_1}^{-l})
(\mu^{(j)}{\alpha^{(j)}_2}^{-l}-\mu^{(k)}{\alpha^{(k)}_2}^{-l})}
{(\mu^{(j)}{\alpha^{(j)}_2}^{-l}-\mu^{(k)}{\alpha^{(k)}_1}^{-l})
(\mu^{(j)}{\alpha^{(j)}_1}^{-l}-\mu^{(k)}{\alpha^{(k)}_2}^{-l})}
$$
after the interaction with $k$-th component. 
If the multi-soliton solution is nonsingular for $t\to-\infty$ and some of 
$\delta_{jk}$ are complex, then the solution became singular after the 
interaction. 
The interaction of solitons of general form is more complicated and changes 
their internal degrees of freedom. 
Since $\mu^{(j)}$ can differ by the sign for fixed set of the Eq.(\ref{a_jk})
roots, Eqs.(\ref{s_1}), as the Toda lattice equations, have the solitons 
propagating on the lattice in both directions. 

In the case of lattice (\ref{s_2}), we have the following formulas of the DT
of direct problem 
\begin{equation}
\tilde\psi_k=\dot\psi_k-\frac{\dot\varphi_k}{\varphi_k}\psi_k,
\label{tpsi_2}
\end{equation}
\begin{equation}
\tilde\xi_k=\frac{\Delta_k}{\varphi_{k-l}},
\label{txi_2}
\end{equation}
\begin{equation}
\tilde\sigma_k=\sigma_k+\log\frac{\varphi_{k-l-1}}{\varphi_{k-l}}. 
\label{ts_2}
\end{equation}
Here $\psi_k$ and $\xi_k$ are solutions of Lax pairs (\ref{Lax_2}) and 
(\ref{Lax_2d}), $\varphi_k$ is solution of (\ref{Lax_2}) with $z=x$, 
$\lambda=\mu$, $\Delta_k$ are defined by Eqs.(\ref{D_1}), where constants 
$\delta_k$ have to satisfy equations
$$
\lambda\delta_{k+l}-\mu\delta_k=\varphi_k\xi_k|_{t=t_0},
$$
$$
x\lambda^2\delta_{k-1}-z\mu^2\delta_k=\left.\left[
\lambda\mu h_k\varphi_k\xi_{k-l-1}+\lambda\rho_{k+l}\varphi_{k+l}\xi_{k-l-1}+
\mu\rho_k\varphi_k\xi_{k-2l-1}\right]\right|_{t=t_0},
$$
(Relations (\ref{D_2}) or (\ref{D_3}) can be also used as the definitions of 
$\Delta_k$.) 
Iterating this DT $N$-times, we obtain the next expression for new solution of 
lattice (\ref{s_2}):
\begin{equation}
\tilde\sigma_k=\sigma_k+
\log\frac{W(\varphi^{(1)}_{k-l-1},...,\varphi^{(N)}_{k-l-1})}
{W(\varphi^{(1)}_{k-l},...,\varphi^{(N)}_{k-l})},
\label{ts_2N}
\end{equation}
where $\varphi^{(j)}_k$ ($j=1,...,N$) are solutions of Lax pair (\ref{Lax_2}) 
with $z=x^{(j)}$, $\lambda=\mu^{(j)}$. 

As for the previous case, one can introduce DT of dual problem and construct 
the binary DT, which gives the following formula for transformed solution 
of lattice (\ref{s_2}): 
\begin{equation}
\tilde\sigma_k=\sigma_k+\log\frac{\Delta_{k-1}}{\Delta_k}.
\label{ts_2b}
\end{equation}
Let us suppose $\lambda=\mu=1$ and consider limit $x=z+\delta z\to z$, 
$\varphi_k=\psi_k+O(\delta z)$. 
Then Eqs.(\ref{ts_2b}) yield
$$
\tilde\sigma_k=\sigma_k+\delta z\,\delta\sigma_k+o(\delta z),
$$
where
$$
\delta\sigma_k=\Delta_{k-1}-\Delta_k
$$
is solution of the linearization of Eqs.(\ref{s_2}). 
After substitution of expansions (\ref{psi_0},\ref{xi_0}) into this formula 
we come to representation 
$$
\delta\sigma_k=\sum_{m=1}^{\infty}\delta\sigma_k^{(m)}z^{ml}. 
$$
The first members of the hierarchy of the infinitesimal symmetries 
$\delta\sigma_k^{(m)}$ of lattice (\ref{s_2}) are 
$$
\delta\sigma_k^{(1)}=h_k,
$$
$$
\delta\sigma_k^{(2)}=\rho_{k+l}+\rho_k-h_k\sum_{m=-l}^lh_{k+m}. 
$$

Solutions $\varphi^{(j)}_k$ of the Lax pair (\ref{Lax_2}) on the zero 
background are represented in the following manner 
\begin{equation}
\chi^{(j)}_k=\sum_{m=1}^{2l+1}D^{(j)}_m{\beta^{(j)}_m}^k
\mbox{e}^{\displaystyle-\nu^{(j)}{\beta^{(j)}_m}^{-l}t},
\label{chi_jk}
\end{equation}
where $\beta^{(j)}_m$ ($m=1,...,2l+1$) are the roots of equations 
\begin{equation}
y^{(j)}={\nu^{(j)}}^2\beta^{(j)}+C{\beta^{(j)}}^{2l+1},
\label{b_jk}
\end{equation}
$D^{(j)}_m$ are constants.
The substitution (\ref{chi_jk}) into (\ref{ts_2N}) gives the multi-soliton 
solution of lattice (\ref{s_2}).
The properties of this solution are similar to ones of lattice (\ref{s_1}).
If all $x^{(j)}$ are different, then the one-soliton component of the 
multi-soliton solution is completely described by parameter $y^{(j)}$ and $2l$ 
constants from set $D^{(j)}_m$ ($m=1,...2l+1$). 
If only two constants $D^{(j)}_m$ for any $j$ are unequal to zero, then an 
interaction of solitons can lead to their shifts.
The expressions for the shifts differ from ones for Eqs.(\ref{s_1}) by the 
notations (compare Eqs.(\ref{phi_jk}) and (\ref{chi_jk})).

\section{Conclusion}

We have obtained the expressions for the first integrals, infinitesimal 
symmetries and the multi-soliton solutions for the sets (\ref{s_1},\ref{s_2}) 
of the integrable lattice equations of the Toda type. 
These equations generalize the Toda lattice on the case of the systems of 
particles interacting with a few neighborhoods and can be considered as the 
reductions of the Nahm equations. 
The evolution of the solutions of the lattices can result in an appearance 
of the new singularities. 
Such the singularities arise no in the case of real-valued solutions of the 
Toda lattice only, when positively defined first integral exists. 
The study of the compatible flows, the symmetries and the discretizations of 
the lattice equations can lead to the new hierarchies of the integrable 
equations \cite{ASh,MS,LM}. 
From the point of view of the quantization of the equations considered, it is 
important to find the hierarchies of the Poisson structures and to include 
them into the $r$--matrix approach \cite{A}. 
This is also significant for proving the integrability of the lattices in the 
periodic case. 

\section{Acknowledgements}
I am grateful to Dr. Marek Czachor for hospitality during my stay at 
Politechnika Gda\'{n}ska and thank Nokia--Poland for financial support.

\vfill
\eject

\begin{thebibliography}{99}
\bibitem{T} M. Toda, J. Phys. Soc. Jpn. 22 (1967) 431.
\bibitem{H} R. Hirota, J. Phys. Soc. Jpn. 35 (1973) 286.
\bibitem{F} H. Flashka, Phys. Rev. B 9 (1974) 1924; Progr. Theor. Phys. 51 
(1974) 703.
\bibitem{M} S.V. Manakov, Zh. Eksp. Theor. Fiz. 67 (1974) 543 (Engl. transl. 
Sov. Phys.-JETP 40 (1975) 269).
\bibitem{In} V. Inozemtsev, Commun. Math. Phys. 121 (1989) 629. 
\bibitem{Per} A.M. Perelomov, {\it Integrable systems of classical mechanics
and Lie algebras} (Nauka, Moskow, 1990). 
\bibitem{Ya} R.I. Yamilov, ''Classification of Toda type scalar lattices,''
in {\it Nonlinear Evolution Equations and Dynamical Systems}, edited by 
V.~Makhankov, I.~Puzynin, and O.~Pashaev (World Scientific, Singapore, 1993), 
pp. 423-431. 
\bibitem{ASh} V.E. Adler, A.B. Shabat, Theor. Math. Phys. 111 (1997) 323; 
112 (1997) 179; V.G.~Marikhin, A.B. Shabat, Theor. Math. Phys. 118 (1999) 217 
(in Russian).
\bibitem{Bl} M. B\l aszak, {\it Multi--Hamiltonian theory of dynamical
systems} (Springer--Verlag, Berlin--Heidelberg, 1998).
\bibitem{Sv} A.K. Svinin, Inverse Problems 17 (2001) 307; J. Phys. A 34 (2001) 
10559; Theor. Math. Phys. 130 (2002) 15 (in Russian).
\bibitem{CCzU} J.L. Cieslinski, M. Czachor, N.V. Ustinov, J. Math. Phys. 44 
(2003) 1763. 
\bibitem{U} N.V. Ustinov, J. Phys. A 35 (2002) 6963. 
\bibitem{AFGZ} H. Aratyn, L.A. Ferreira, J.F. Gomes, A.H. Zimmermann, ''Toda 
and Volterra lattice equations from discrete symmetries of KP hierarchies,'' 
{\it Preprint} hep-th/9307147. 
\bibitem{LShYa} A.N. Leznov, A.B. Shabat, R.I. Yamilov, Phys. Lett. A 174 
(1993) 397. 
\bibitem{Marsh} A. Marshakov, {\it Seiberg--Witten theory and integrable 
systems} (World Scientific, Singapore, 1999). 
\bibitem{Nahm} W. Nahm, Phys. Lett. B90 (1980) 413; ''The construction of all 
self--dual multimonopoles by the ADHM method,'' in {\it Monopoles in Quantum 
Field Theories}, edited by N. Craigie, P. Goddard, and W. Nahm (World 
Scientific, Singapore, 1982), pp. 440
\bibitem{Hitchin} N.J. Hitchin, Commun. Math. Phys. 89 (1983) 145. 
\bibitem{D} D. Diaconescu, Nucl. Phys. B 503 (1997) 220 
\bibitem{GGM} A. Gorsky, S. Gukov, A. Mironov, Nucl. Phys. B 517 (1998) 409 
\bibitem{GKMMM} A. Gorsky, I. Krichever, A. Marshakov, A. Mironov, A. Morozov, 
Phys. Lett. B 355 (1995) 466; A. Gorsky, A. Marshakov, Phys. Lett. B 374 
(1996) 218 
\bibitem{MW} E. Martinec, N. Warner, Nucl. Phys. B 459 (1996) 97 
\bibitem{IM} H. Itoyama, A. Morozov, Nucl. Phys. B 491 (1997) 529 
\bibitem{MaSa} V.B. Matveev, M.A. Salle, {\it Darboux Transformations 
and Solitons\/} (Springer--Verlag, Berlin--Heidelberg, 1991).
\bibitem{B1} O.I. Bogoyavlensky, Phys. Lett. A 134 (1988) 34.
\bibitem{BCh} A.A. Belov, K.D. Chaltikian, Phys. Lett. B 309 (1993) 268.
\bibitem{B2} O.I. Bogoyavlenskii, Commun. Math. Phys. 51 (1976) 201.
\bibitem{AS} M.J. Ablowitz, H. Segur, {\it Solitons and the Inverse Scattering 
Transform\/} (SIAM, Philadelphia, 1981).
\bibitem{MS} A.V. Mikhailov, V.V. Sokolov, Theor. Math. Phys. 122 (2000) 72 
(in Russian).
\bibitem{LM} D. Levi, L. Martina,  J. Phys. A 34 (2001) 10357.
\bibitem{A} V.I. Arnold (ed.), {\it Encyclopaedia of Mathematical Sciences. 
Vol.3. Dynamical Systems III\/} (Springer--Verlag, Berlin--Heidelberg, 1993).
\end{thebibliography}
\end{document}